\documentclass[aps,prl,preprint,groupedaddress,showpacs]{revtex4}
\usepackage{graphicx}
\begin{document}

\title{Comment on 'Quantum Coherence between High Spin Superposition States of Single Molecule
Magnet Ni$_4$'}

\author{W. Wernsdorfer}

\affiliation{
Laboratoire Louis N\'eel, associ\'e \`a l'UJF, CNRS, BP 166, 38042 Grenoble Cedex 9, France
}

\date{20 May 2004}

\begin{abstract}
In a recent paper, http://xxx.lanl.gov/abs/cond-mat/0405331 (Ref. 1),
del Barco et al. reported experimental studies on a Ni$_4$
molecular system.
They used an experimental method (combining microwave spectroscopy with high 
sensitivity magnetic measurements) that we have introduced before,
http://xxx.lanl.gov/abs/cond-mat/0404410 (see also Ref. 2 and 3).
Among other things, our technique allows us to monitor 
spin-state populations in the presence of microwave magnetic fields.
Absorption line-widths give rough 'upper bounds' on the decoherence rate
similar to 'standard' high frequency electron paramagnetic resonance 
(HF-EPR) techniques.
In the case of quasi continuous radiation
our technique does {\bf NOT} 
give directly the spin-lattice relaxation time $T_1$.
For measurements like those of del Barco et al., it simply gives access to 
the phonon-bottleneck time, a parameter that is many orders of magnitude longer 
than the spin-lattice relaxation time. Any conclusion concerning
quantum coherence is preliminary.
\end{abstract}

\pacs{75.45.+j, 75.60.Ej, 75.50.Xx, 42.50.Fx}
\maketitle

It is widely accepted that single-molecule magnets (SMMs)
are interesting new model systems to study quantum dynamics.
With respect to diluted paramagnetic ion systems, 
the extraordinary tools of organic and coordination 
chemistry allow the design of new (supra)molecular 
systems with promising properties~\cite{Christou00}.
SMMs straddle the interface 
between classical and quantum mechanical behavior because they also 
display quantum tunneling of magnetization
~\cite{Novak95,Friedman96,Thomas96,Sangregorio97,Hill98,Aubin98,Soler04,Tasiopoulos04} 
and quantum phase interference~\cite{Garg93,WW_Science99}.

We reported recently~\cite{WW_0404410,remark1} a technique allowing us to monitor 
spin-state populations in the presence of microwave magnetic fields.
Absorption line-widths give rough 'upper bounds' on the decoherence rate
similar to 'standard' high frequency electron paramagnetic resonance 
(HF-EPR) studies~\cite{Barra97,Hill98,Hill_Science03}.
The advantages of our technique with respect to
pulsed EPR techniques
involves the possibility to perform time-resolved experiments 
(below 1 ns)~\cite{ChiorescuScience03} 
on submicrometer sizes samples (about 1000 spins)~\cite{Jamet01a} 
at low temperature (below 100 mK).
Our first results on the V$_{15}$ system open the way 
for time-resolved observations
of quantum superposition of spin-up and spin-down states in SMMs. 
Other results obtained in similar systems but with large spins
concern for example EPR measurements~\cite{Hill98}, 
resonant photon-assisted tunneling 
in a Fe$_8$ SMM~\cite{Sorace03} and non-resonant microwave absorption
in a Mn$_{12}$ SMM~\cite{Amigo03}.

\begin{figure}
\begin{center}
\includegraphics[width=.48\textwidth]{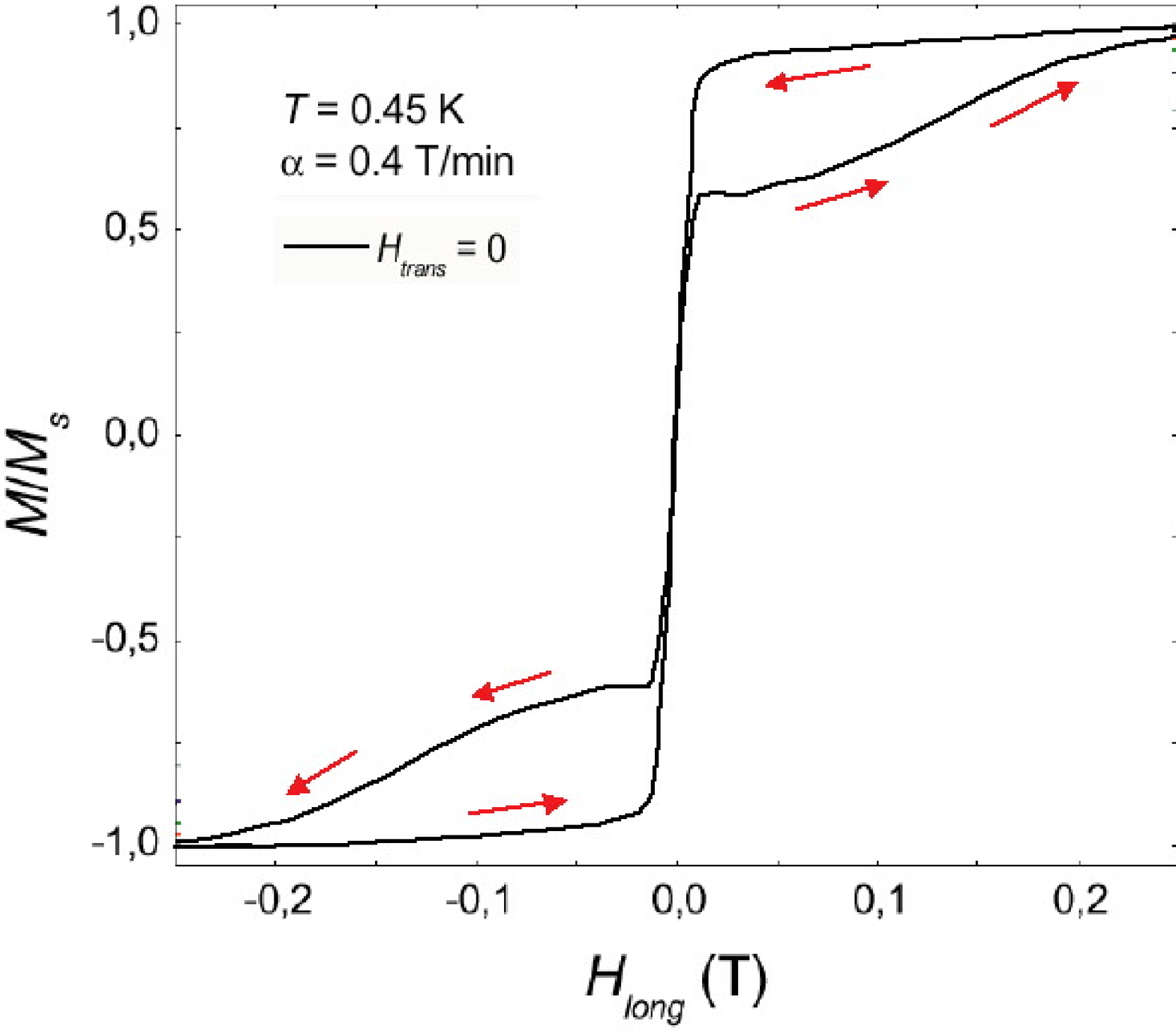}
\includegraphics[width=.48\textwidth]{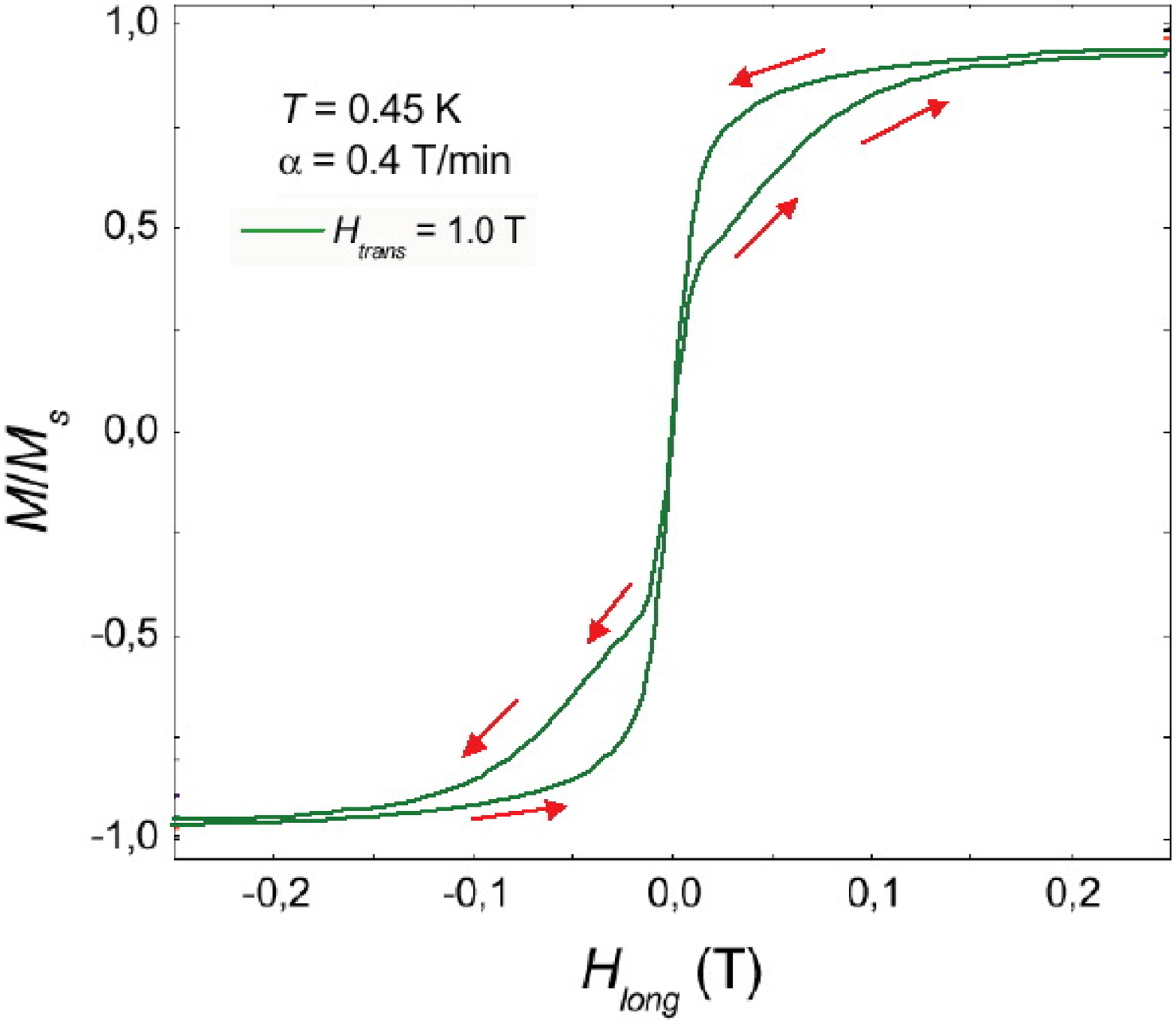}
\includegraphics[width=.48\textwidth]{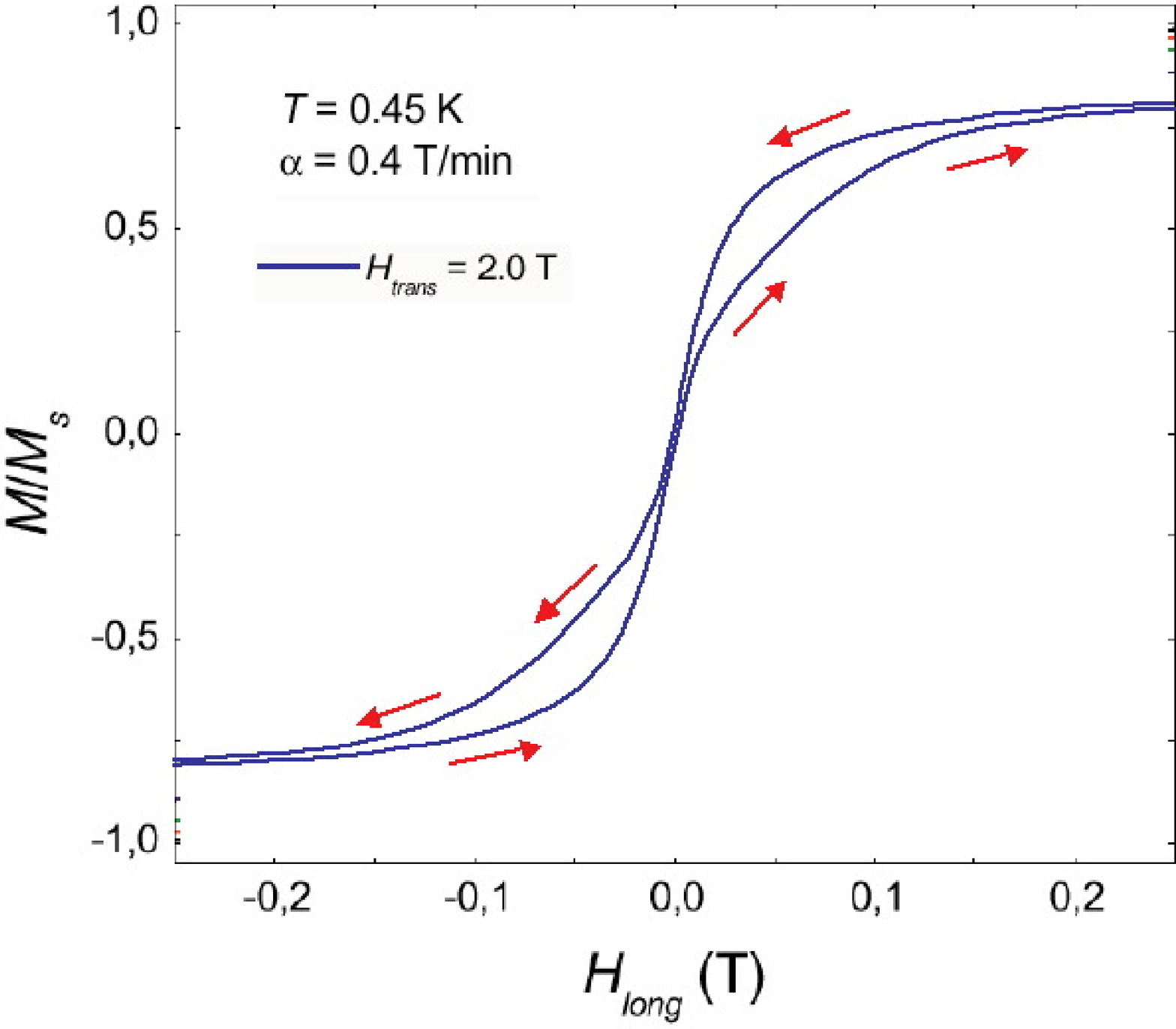}
\caption{(color)  (a) Hysteresis loop measurements
for a single crystal of Ni$_{4}$ at several
transverse fields. The data were scanned from
Fig. 1 in cond-mat/0405541. The back sweeps were
obtained by a symmetry operation. Such loops are
typical for a  phonon-bottleneck effect~\cite{ChiorescuV15PRL00,Schenker,Waldmann}.
Note that the field sweep rate is very small. The opening at this
sweep rate corresponds to relaxation rates of a few tens of seconds,
in agreement with Fig. 3 in Ref. 1. 
}
\label{fig2}
\end{center}
\end{figure}

In a recent paper
del Barco et al. reported experimental studies on a Ni$_4$
molecular system~\cite{Barco_0405331} similar to our 
study~\cite{WW_0404410,remark1}.
However, they replaced the micro-SQUID by micro-Hall sensors.
The latter has a lower sensitivity and it has not yet been
shown that time-resolved experiments in the nanosecond range
are possible.
In the case of quasi continuous radiation
our technique does {\bf NOT} 
give directly the spin-lattice relaxation time $T_1$.
For measurements like those of del Barco et al.~\cite{Barco_0405331}, 
it simply gives access to 
the phonon-bottleneck time~\cite{WW_0404410,remark1}, 
a parameter that is many orders of magnitude longer 
than the spin-lattice relaxation time.

This comment recalls briefly the phonon-bottleneck effect that
was first studied in diluted paramagnetic ion systems~\cite{Abragam70}.
In the field of molecular systems, this effect was rediscovered
~\cite{ChiorescuV15PRL00,Schenker,Waldmann}
and it plays a certain role in all currently available molecular systems. 
The energy exchange between the spin system and the cryostat (heat bath)
goes via the phonon lattice modes of the crystal. 
A phonon-bottleneck occurs as soon as the heat capacity
of the phonons is much smaller than that of the spins.
The energy $\Delta_H$ is transferred from the spins to only 
those phonon modes with the energy $h \nu = \Delta_H$
(within the resonance line-width). 
Because the number of such lattice modes is
much smaller than the number of spins, 
the energy transfer is very difficult, leading
to a phenomenon known as the phonon-bottleneck~\cite{Abragam70}.

The phonon-bottleneck time $\tau_{\rm ph}$ depends on
many experimental conditions: spin value, magnetic 
anisotropy of the spins, temperature, applied field,  
crystal size and shape, thermalization, sample holder, etc.
Generally, for molecular systems below a few kelvin and for
fields larger than a few tens of mT, $\tau_{\rm ph}$ ranges
from a few seconds to 1000 s~\cite{Chiorescu03}.
The temperature dependence of $\tau_{\rm ph}$ does not 
follow an Arrhenius law~\cite{ChiorescuV15PRL00,Chiorescu03}.

The relaxation rates reported by del Barco et al.
~\cite{Barco_0405331} on a Ni$_4$
molecular system are typical for a phonon-bottleneck
and had been measured by micro-SQUID measurements on 
the same system (unpublished, see also Fig.1). Each pulse of
microwaves excites spins. It will take then
a time $\tau_{\rm ph}$ in order to transfer
the energy from the spin system to the cryostat (heat bath).
This heat transfer is 'slowly' because of the
small heat capacity.
The relaxation times are therefore not
remarkably long and the increase from 8 to 20 s as the field
increases is {\bf not} contrary to general
ideas that the relaxation time should decrease with frequency. 
In the phonon-bottleneck regime, such an increase
is expected until the field reaches an energy splitting
of a few kelvin. More details have been presented (Ref. 3)
and will be published elsewhere.

Finally, we mention
that this system does not block at zero field because
of fast tunnel dynamics (even without transverse field). 
Below about 0.3 K, the system orders because of a small
antiferromagnetic intermolecular exchange coupling.
Both the conclusion that $\tau_2 \sim \tau_{\phi}$ 
and all statements concerning the observation of quantum 
coherence are not yet demonstrated.
The RF-power estimations in Ref. 1 should take into account
the phonon-bottleneck effect.

\newpage


\end{document}